\input harvmac

\def\Title#1#2{\rightline{#1}\ifx\answ\bigans\nopagenumbers\pageno0
\else\pageno1\vskip.5in\fi \centerline{\titlefont #2}\vskip .3in}

\font\caps=cmcsc10

\noblackbox
\parskip=1.5mm


\def\npb#1#2#3{{\it Nucl. Phys.} {\bf B#1} (#2) #3 }
\def\plb#1#2#3{{\it Phys. Lett.} {\bf B#1} (#2) #3 }
\def\prd#1#2#3{{\it Phys. Rev. } {\bf D#1} (#2) #3 }

\def\ijmpa#1#2#3{{\it Int. J. Mod. Phys.} {\bf A#1} (#2) #3 }

\def\bb#1{{\tt hep-th/#1}}


\def\CL{{\cal L}}


\def\dj{\hbox{d\kern-0.347em \vrule width 0.3em height 1.252ex depth
-1.21ex \kern 0.051em}}

\def\half{{1\over 2}\,}

\def\Tr{{\rm Tr\,}}

\def\pt{\partial}

\def\ep{\varepsilon}
\def\zb{\overline z}

\lref\rgiv{A. Giveon and D. Kutasov, {\it ``Brane Dynamics and Gauge Theory'',}
\bb{9802067.}}
\lref\regk{S. Elitzur, A. Giveon and D. Kutasov,
\plb{400}{1997}{269,} \bb{9702014.}}
\lref\regkrs{S. Elitzur, A. Giveon, D. Kutasov, E. Rabinovici and 
A. Schwimmer, \npb{505}{1997}{202,} \bb{9704104.}}
\lref\rhw{A. Hanany and E. Witten, \npb{492}{1997}{152,} \bb{9611230.}}
\lref\rusdo{J.L.F. Barb\'on and A. Pasquinucci, {\it ``D0-branes,
Constrained Instantons and D=4 Super Yang--Mills Theories'',}
\bb{9708041,} to appear in {\it Nucl. Phys.} {\bf B};
{\it ``D0-Branes as Instantons in D=4 Super Yang-Mills Theories'',}
\bb{9712135.}}
\lref\rwittenM{E. Witten, \npb{500}{1997}{3,} \bb{9703166.}}
\lref\rkahler{J. de Boer, K. Hori, H. Ooguri and Y. Oz,
{\it ``Kahler Potential and Higher Derivative Terms from M Theory Fivebrane'',}
\bb{9711143.}}
\lref\rwitqcd{E. Witten, \npb{507}{1997}{658,} \bb{9706109.}}
\lref\rbarb{J.L.F. Barb\'on, \plb{402}{1997}{59,} \bb{9703051.}}
\lref\rthetus{J.L.F. Barb\'on and A. Pasquinucci,
{\it ``Softly Broken MQCD and the Theta Angle'',} \bb{9711030,}
to appear in {\it Phys. Lett.} {\bf B}.} 
\lref\rsoft{O. Aharony, J. Sonnenschein, M.E. Peskin and S. Yankielowicz,
\prd{52}{1995}{6157,}  \bb{9507013.}}
\lref\rbst{A. Brandhuber, J. Sonnenschein, S. Theisen and S. Yankielowicz,
\npb{502}{1997}{125,} \bb{9704044.}}
\lref\rmberk{K. Hori, H. Ooguri and Y. Oz, {\it Adv. Theor. Math. Phys.} 
{\bf 1} (1998) 1, \bb{9706082.}}
\lref\rbikap{A. Brandhuber, N. Itzhaki, V. Kaplunovsky, J. Sonnenschein and
 S. Yankielowicz, \plb{410}{1997}{27,} \bb{9706127.}}
\lref\rozdb{J. de Boer and Y. Oz, \npb{511}{1998}{155,} \bb{9708044.}}
\lref\rads{I. Affleck, M. Dine and N. Seiberg, \npb{241}{1984}{493.}}
\lref\rlastev{N. Evans, {\it ``Quark Condensates in Non-supersymmetric MQCD'',}
\bb{9801159.}}
\lref\revans{N. Evans and M. Schwetz,
{\it ``The Field Theory of Non-Supersymmetric Brane Configurations'',}
\bb{9708122.}}
\lref\rtheta{N. Evans, S. Hsu and M. Schwetz, \npb{484}{1997}{124,}
\bb{9608135,~} \plb{404}{1997}{77,} \bb{9703197\semi}
K. Konishi and M. Di Pierro, \plb{388}{1996}{90,} \bb{9605178\semi}
K. Konishi, \plb{392}{1997}{101,} \bb{9609021.}}
\lref\ryale{N. Evans, S. Hsu and M. Schwetz, \plb{355}{1995}{475,}
\bb{9503186\semi} N. Evans, S. Hsu, M. Schwetz and S.B. Selipsky,
\npb{456}{1995}{205,} \bb{9508002.} }
\lref\rlag{L. Alvarez-Gaum\'e, J. Distler, C. Kounnas and
M. Mari\~no, \ijmpa{11}{1996}{4745,} \bb{9604004\semi}
L. Alvarez-Gaum\'e and M. Mari\~no, \ijmpa{12}{1997}{975,} \bb{9606191\semi}
L. Alvarez-Gaum\'e, M. Mari\~no and F. Zamora, \ijmpa{13}{1998}{403,} 
\bb{9703072;}
{\it ``Softly Broken N=2 QCD with Massive Quark Hypermultiplets, II'',}
\bb{9707017.} }



\line{\hfill CERN-TH/98-112}
\line{\hfill KUL-TF-98/18}
\line{\hfill {\tt hep-th/9804029}}
\vskip 1.2cm

\Title{\vbox{\baselineskip 12pt\hbox{}
 }}
{\vbox {\centerline{A Note on Softly Broken MQCD}
}}

\vskip0.6cm

\centerline{$\quad$ {\caps J. L. F. Barb\'on~$^1$ and A. Pasquinucci~$^2$
 }}
\vskip0.8cm

\centerline{{\sl $^1$ Theory Division, CERN}}
\centerline{{\sl 1211 Geneva 23, Switzerland}}
\centerline{{\tt barbon@mail.cern.ch}}

\vskip0.3cm

\centerline{{\sl  $^2$ Instituut voor theoretische fysica, K.U.\
Leuven}}
\centerline{{\sl  Celestijnenlaan 200D }}
\centerline{{\sl  B-3001 Leuven, Belgium  }}

\vskip 1.0in

\noindent{\bf Abstract:} 
We consider generic MQCD configurations with matter described by 
semi-infinte D4-branes and softly broken supersymmetry. 
We show that the matter sector does not introduce supersymmetry 
breaking parameters so that the most relevant supersymmetry
breaking operator at low energies is the gaugino mass term.
By studying the run-away properties of these models in the
decoupling limit of the adjoint matter, we argue that these softly
broken MQCD configurations fail to capture the infrared physics of QCD
at scales below the gaugino mass scale. 

%
%
%


\Date{March 1998}


%
\newsec{Introduction and Conclusions}
The modeling of gauge dynamics via brane configurations has
provided a nice geometric picture of some very non-trivial
infrared phenomena of gauge theories, like the dynamical
generation of superpotentials, confinement and gaugino condensation
in $N=1$ theories, and the complete Seiberg-Witten effective
action in $N=2$ theories.  A recent review with an extensive
list of references is \refs\rgiv.  The systematics  of the
engineering
of gauge theories in the world-volume of  complicated
brane configurations was initially developed
in \refs\rhw. The advantage
of the brane set up at weak coupling is that the structure of
the classical moduli space of vacua is given a geometric interpretation
in terms of the rules of brane dynamics in, say, weakly coupled
type IIA string theory. Quantum corrections in the weak coupling
regime include bending effects of the branes \refs\rwittenM, 
D-instanton corrections \refs\rusdo, and other effects which
comprise a set of ``force rules"    
of brane dynamics \refs\regkrs. An efficient way of summing
up all these corrections was proposed by Witten in \refs\rwittenM. 
Roughly speaking, one takes the IIA string theory to strong coupling,
and tries to calculate non-perturbative effects in terms of the
perturbation theory of the dual theory. In this case the dual 
theory is M-theory, and one is confined to the infrared domain,
where it admits a description in terms of eleven dimensional 
supergravity. At strong coupling, the quantum corrections
like D-instanton effects are summed up into the background
geometry of the M-theory solitonic configuration; a 
single  twisted M-theory five-brane in the simplest examples.   

{}From the point of view of the low energy field theory, changing  
the string coupling  keeping the gauge coupling
fixed, corresponds to turning on a tower of perturbatively
irrelevant operators. These operators decouple in the infrared
as long as we are in the weak coupling regime, but in general
there is no guarantee that they continue to do so in the strong
coupling regime. As a result, one discovers that the supergravity
description in terms of the M5-brane (the MQCD regime) depends
explicitly on the radius of the new dimension $R$, in addition
to the low energy scale of gauge theory $\Lambda_{QCD}$. Holomorphic
observables (protected by supersymmetry) exactly match the gauge
theory ones, but the $R$ dependence shows up in higher derivative
interactions \refs\rkahler\ and in massive states in theories
with a mass gap \refs\rwitqcd.

Soft supersymmetry breaking is introduced geometrically by
rotations of the branes in the type IIA picture at weak
coupling \refs{\rbarb,\regk,\rbst}, or by rotations of
the asymptotic boundary conditions of the M5-brane in the
M-theory picture \refs\rwitqcd.~\foot{Discussions of
various issues concerning soft supersymmetry breaking in SQCD and
MQCD can be found in refs.\ 
\refs{\ryale,\rsoft,\rlag,\rtheta,\revans}.}
In the light of the previous
comments, it is to be expected that the loss of supersymmetry
constraints makes the matching of MQCD and true QCD physics
less succesful. Still, MQCD in itself is an interesting theory
whose supersymmetry breaking dynamics is worth study. In addition,
it has been shown in \refs\rthetus\ that some non-trivial vacuum
physics, like level crossing phenomena, can be detected geometrically
in MQCD. 

In this note we add some comments on the interpretation of
non-supersymmetric brane configurations. First, we  show 
that, in a general class of models, many of the supersymmetry
breaking deformations naively present in the type IIA brane 
picture (rotation angles), cannot be realized smoothly in
the M-theory limit. The requirement that the five-brane configurations
are smooth, restricts some of the possible rotation angles. In
particular, in models with semi-infinite D4-branes representing
hypermultiplet couplings, the asymptotic orientation of the    
D4-branes in eleven dimensional space does not introduce any
new supersymmetry breaking couplings. Therefore, in this
class of models all soft parameters are associated to rotations
of the NS-branes of type IIA, or in field theory terms, to 
soft parameters in the $N=2$ vector multiplet sector. The most
relevant 
supersymmetry breaking operator at low energies is then the 
gaugino mass term $\CL_{\rm soft} = m_{\lambda} \Tr \lambda 
\lambda$.  It is important to emphasize that this result
applies to models with supersymmetry breaking rotations 
restricted to NS branes and semi-infinite D4-branes. In principle,
models with matter described by D6-branes could have a 
larger parameter space and
non trivial rotations of the D6-branes could be important in
generating soft parameters in the flavour sector. However,
experience with the $N=2$ and $N=1$ models shows that qualitative
features of the Higgs phases like the existence of run-away
vacua, can be established at the level of the eigenvalues of
the meson matrices, which are correctly captured by semi-infinite
D4-branes.       

The interpretation of this result depends on the extent to
which the brane configuration captures the infrared behavior
of the theory. On general grounds, at scales below the gaugino
mass, radiative corrections would induce a mass for the squarks
as well, supressed only by some powers of the coupling. The fact
that MQCD represents the strong coupling regime makes it possible
for this squark mass to be incorporated into the brane geometry,
even if it does not appear in the asymptotic boundary conditions
of the brane configuration. In other words, we would like to
know if the brane contains information down to the deep
infrared domain, at energies $E<m_{\lambda}$, or only   
captures QCD physics down to an intermediate scale 
$m_{\lambda} < E < \Lambda_{QCD}$. 

We investigate this question by examining the run-away 
vacua of $N=1$ super QCD for $N_f < N_c$, and the possibility
of vacuum stabilization by a squark mass. In these models,
say the $N_f = N_c -1$ case, the Affleck-Dine-Seiberg 
superpotential \rads
$$
W_{ADS} = {\Lambda_1^{2N_c +1} \over {\rm det}({\widetilde Q} Q)} 
$$
induces a potential that pushes the vacuum expectation value
of the squarks to infinity. If a small squark mass would be
added to the bare lagrangian or generated dynamically at some
scale, it would stop the run-away.
 As pointed out in ref.\ \refs\rsoft, such vacua are physically
very interesting, as one can argue for chiral symmetry breaking
with the standard vectorlike pattern of QCD.   

A geometric criterion for
run-away in the brane configurations of M-theory was introduced
in ref.\  \refs\rmberk\ (see also ref.\ \refs\rbikap). 
Here we study the impact of supersymmetry
breaking rotations on the runway behavior of such configurations. 
We analyze the renormalization group flows of the soft breaking
parameters in the limit where one decouples the adjoint matter
of the $N=2$ vector multiplet. The result is that run-away
behavior is not affected by soft supersymmetry breaking, and
the models with $N_f <N_c$ show no stable vacuum after a gaugino
mass is introduced. 

This result means that these brane configurations fail to
capture the deep infrared region of the field theory, even at
the level of the vacuum structure.  In order to do so, one
presumably would need to incorporate non trivial backgrounds,
i.e.\ D6-branes, in the determination of soft parameters
of the flavour sector.

\newsec{Regularity constraints on fivebrane configurations}

We consider general IIA  brane configurations combining NS-branes and
finite or semi-infinite D4-branes, the latter representing only
flavour degrees of freedom.   These models support product gauge
groups, each factor with a rank given by the number of D4-branes
suspended between contiguous NS-branes.  
Each of the NS-branes could have a high multiplicity,
allowing for the introduction of non-trivial Landau-Ginzburg interactions
\refs\rozdb.
We shall assume that a general configuration with softly broken 
supersymmetry has been
obtained by asymptotic rotation of the branes involved from a completely
degenerate configuration of the Coulomb branch, i.e.\ the complete
M-theory  fivebrane has the structure  $M_4 \times {\bf C} (z_{\alpha})$,
with $M_4$ the standard $3+1$ dimensional Minkowski space, and ${\bf C}  
(z_{\alpha})$ a punctured complex plane. 
The interpretation of the  punctures
is that the region around them is mapped to asymptotic infinity in the
M-theory space-time $R^{1,9} \times S^1_R$,
approaching the type IIA configuration. We denote the embedding coordinates
by $X(z,\zb)$, with $X=({\vec X}, X^{10})$, and 
${\vec X} =(X^4, X^5, X^6, X^7, X^8, X^9)$. 
The background metric in the relevant seven-dimensional space is 
\eqn\metric
{ds^2 =  (dX)^2 = (d{\vec X})^2 +  (dX^{10})^2
\,,  }
so that the minimal area equations for the embedded surface $X(z,\zb)$
are equivalent to the harmonic condition $\pt_z  \pt_{\zb}  X(z,\zb)=0$, 
plus the Virasoro constraints
$T_{zz} =  (\pt_z  X)^2 =0$.

The general ansatz compatible with the  harmonicity  is 
$X(z,\zb) = \sum_{\alpha} \half (X_{\alpha}(z) + {\rm c.c.})$, and
\eqn\genan{
 X_{\alpha} (z) =     x_{\alpha} +
 P_{\alpha} \,{\rm log}\,(z-z_{\alpha}) + \sum_{m \neq 0}
 C_{\alpha,m} \, (z-z_{\alpha})^{m}    }
and the asymptotic conditions at the punctures are specified by
the coefficients in this expansion. We can restrict them by
physical considerations. First, if we define $n_{\alpha}$
to be the wrapping number at $z_{\alpha}$
 of the M5 around the circle $X^{10}$, we can choose the normalization
of the $z$ coordinate such that the monodromy  under $(z-z_{\alpha})
\rightarrow (z-z_{\alpha}) e^{2\pi i}$
 is given by $X^{10} \rightarrow X^{10} + 2\pi R  n_{\alpha}$. Then
using the gauge freedom to shift $X(z,\zb)$ by a single harmonic
function we can reduce   $X^{10}$ to the simplest form compatible
with the monodromy:
\eqn\exten
{X^{10}_{\alpha}(z) = -i R \, n_{\alpha} \,{\rm log}\,(z-z_{\alpha})  }
Since the wrapping number around $X^{10}$ is basically the D4-brane
number in this construction, the integers $n_{\alpha}$ are interpreted
as the number of D4-branes from the ``left" minus the number from the
``right" when the puncture $z_{\alpha} =z_a $ is associated to an NS-brane
in the type IIA limit. In the following, it will be convenient to
distinguish the index $\alpha = (a,k)$, between the punctures
associated to asymptotic NS-branes, with index $a$, and the
punctures associated to semi-infinite D4-branes, with index $k$.
For the latter case,  
the number  $n_{k}$ is interpreted simply as the number of
D4-branes sticking out of the puncture.

Now, in the IIA limit, an asymptotic NS-brane with multiplicity $h_\alpha$
is a plane in ${\vec X}$-space, with an affine representation
\eqn\plane
{{\vec X} \sim {\rm Re}\,({\vec C} \, y).}
Here, ${\vec C}$ is a complex vector whose real and imaginary parts
generate the plane, and $y$ is a complex variable encoding the
affine parameters. The multiplicity is obtained by setting 
$(z-z_{\alpha})^{h_\alpha} =1/y$, so that the $z$-plane around $z_{\alpha}$ 
covers the $y$ plane $h_\alpha$ times.
Therefore the remaining non-trivial embedding function is given by
\eqn\ans{
{\vec X}_{\alpha} (z) =  {\vec x}_{\alpha} +
{\vec P}_{\alpha} \,{\rm log}\,(z-z_{\alpha}) + 
\sum_{n\neq 0}
{\vec C}_{\alpha,n} \, (z-z_{\alpha})^n }
where ${\vec C}_{\alpha,n}$ vanishes for $n<-h_\alpha$. The logarithmic
terms in the $z_a$ punctures, corresponding to NS-branes, reflect
the ``bending effect" discussed in \refs\rwittenM. The D4-branes
ending on the NS-branes behave as sources of codimension two in
the NS-brane world-volume, with charges ${\vec P}_a$.  
For a semi-infinite D4-brane all complex vectors ${\vec C}_{k,n}$ vanish
and only the logarithmic terms remain, because the semi-infinite
D4-branes do not have a plane at infinity in the ten-dimensional
internal space. Indeed, taking ${\vec C}_{k,n}$ real and non-vanishing
would not describe a plane, and moreover the monodromy around the 
$X^{10}$ direction would not act trivially on the resulting 
semi-infinite D4-brane. So all complex vectors ${\vec C}_{k,n}$ vanish
for semi-infinite D4-branes.

Finally, the most important constraint is obtained by requiring
that all the ``momenta" ${\vec P}_{\alpha}$ 
in the non-compact directions are real,
i.e.\ that the monodromy under $(z-z_{\alpha}) \rightarrow
(z-z_{\alpha}) e^{2\pi i}$ only affects $X^{10}$ and leaves
${\vec X}$ invariant. This is a condition for the embedding
to be smooth. 

Now it is straightforward to  write explicitly the 
Virasoro constraints in terms of the vectors 
$B_{\alpha,n} = n C_{\alpha,n} + \delta_{n,0} P_{\alpha}$, which
are just the Fourier components of $\pt_z X_{\alpha} (z)$. They
read:
\eqn\conds{\eqalign{ & B_{\alpha,n} \cdot B_{\beta,m} =0 \;,\;\;\alpha
\neq \beta \cr
& \sum_m B_{\alpha,m} \cdot B_{\alpha, n-m} =0,  }}
or, in terms of the previous variables:
\eqn\condss{\eqalign{
& {\vec C}_{\alpha,n} \cdot {\vec C}_{\beta, m} = {\vec C}_{\alpha, n} 
\cdot {\vec P}_{\beta} = {\vec P}_{\alpha} \cdot {\vec P}_{\beta} -
R^2 n_{\alpha} n_{\beta} =0\;,\;\;\;\;\;\alpha\neq\beta \cr
&\delta_{n,0}\,
\left({\vec P}_{\alpha}^2 - R^2 n_{\alpha}^2 \right)
 +2n{\vec P}_{\alpha} \cdot
{\vec C}_{\alpha,n} + \sum_{m} m(n-m){\vec C}_{\alpha,m} \cdot
{\vec C}_{\alpha, n-m} =0}}
Our main result stems now from the fact that semi-infinite D4-branes
do not have a plane at infinity in ${\vec X}$-space, so that
${\vec C}_{k,n} =0$, and the last equation simplifies dramatically
for the flavour punctures: ${\vec P}_k^2 = R^2 n_k^2$. Combining
this equation with
${\vec P}_k \cdot {\vec P}_{\ell} = R^2 n_k n_{\ell}$,
we obtain that all ${\vec P}_k$ vectors must be collinear.
Using the $SO(6)$ isometries of the ${\vec X}$-space
we can reduce one of the ${\vec C}$ to canonical form, say
${\vec C}_{\infty,1} = (1, -i, 0,0,0,0)$.~\foot{We denote by 
${\vec C}_{\infty,1}$ the coefficient of the highest power in $z$, 
i.e.\ the leading term in the
$\vert z \vert \rightarrow \infty$ limit.}
Then, the previous equations, together with
the real character of $P_{\alpha}$, imply
that no momentum vector ${\vec P}_k$ lies in the $(X^4, X^5)$ plane,
and therefore the semi-infinte D4-branes can be conventionally taken to
point in the $X^6$ direction: ${\vec P}_k = (0,0,\eta\,R\,n_k, 0,0,0)$,
with $\eta = \pm 1$, and there is still a full $SO(3)$ isometry
remaining in the $(X^7, X^8, X^9)$ directions.
The only remaining general rule is that all the asymptotic
NS-planes are orthogonal to the $X^6$ direction (i.e.\ $C^6_{a,n}=0$),
due to the equation ${\vec P}_k \cdot {\vec C}_{a,n} =0$.

The fact that all ${\vec P}_k$ vectors are parallel and conventionally
aligned along the $X^6$ direction, means
that there are no new supersymmetry breaking couplings induced by
the flavour sector.
In other words, the flavour sector, represented here by semi-infinite
D4-branes, is overconstrained and it does not introduce extra soft
supersymmetry breaking parameters. Thus, in these models,
breaking supersymmetry by making a generic rotation of the branes
corresponds, in field theory terms, to introduce soft supersymmetry
breaking parameters in the $N=2$ vector multiplet sector only. At 
sufficient low energies, the most relevant supersymmetry breaking
mechanism is that of giving a tree-level mass to the gauginos.

On the other hand, the fact that all flavour D4-branes
point in a given direction means that their relative distances
define vacuum parameters of the model. This is analogous to the
fact that the gauge coupling is a modulus of the theory for a 
brane configuration representing a conformally invariant fixed
point, as the NS-branes ``bend" in a parallel fashion asymptotically
 \refs\rwittenM. In the supersymmetric limit these parameters
are associated to the eigenvalues of the meson matrix \refs\rmberk,
and this is therefore the natural interpretation as well upon
supersymmetry breaking. This allows us to study the run-away
phenomenon after supersymmetry breaking, just by looking
at the positions of the flavour D4-branes in the transverse
hyperplane.

It is worth to make a comment on the supersymmetric case. Supersymmetry
requires that the embedding is holomorphic with respect to some complex
structure of a six-dimensional internal (sub-) manifold. Our (arbitrary)
choices of orientation fix this almost completely. Indeed we should
set $v=X^4+iX^5$ and the direction in which we have oriented the
semi-infinite D4-branes implies that $s=(X^6\pm iX^{10})/R$, or better
$t=\exp(-s)$. Choosing the plus sign, we obtain that supersymmetry,
with our choices of axes, requires $P^6_\alpha=Rn_\alpha$. When we
have $N=1$ supersymmetry, we set $w=X^8+iX^9$, $X^7=0$, whereas with $N=2$
also $w=0$. 

Since the supersymmetry soft breaking does not depend on the matter
content of the curve, we can restrict ourselves to consider the simplest
models without losing generality. We thus consider the curve describing
(in type-IIA language) two NS-branes with $N_c$ D4-branes suspended 
between them and $N_f$ semi-infinite D4-branes. By convention we set
the two NS-branes at $z=0$ and $\vert z\vert=\infty$ and we keep fixed
the NS-brane at $\vert z\vert=\infty$ (corresponding to the vector
${\vec C}_{\infty,1}$). $r$ of the $N_f$ D4-branes are at $z=z_+$ and
$N_f-r$ at $z=z_-$. The curve is
\eqn\ourcurve{\eqalign{ & v=  v_0 - m_f + z + {\eta\over z}  + 
{\epsilon\over \bar{z}} \ ,\qquad\qquad\qquad\quad w = {\zeta\over z} + 
{\bar\lambda\over \bar{z}}\ , \cr
& t= {z^{N_c}\over (z-z_+)^r (z-z_-)^{N_f-r} }\ , \qquad\qquad\qquad 
X^7= 2\sqrt{\epsilon} \log\left\vert{z \over \Lambda_2} \right\vert \cr }}
where $\eta\epsilon +\zeta\lambda=0$. Here we introduced new 
complex parameters $\eta$, $\zeta$,  and $\lambda$, and the real 
parameter $\epsilon$; these parameters are 
combinations of the $C$s and of the $P$s.
The $x_\alpha$ (i.e.\ $v_0$) can be chosen
so that $v\vert_{z=z_\pm, \epsilon=0}=-m_f$.

Notice that the curve is invariant under $\zeta\leftrightarrow\lambda$ with
$w\leftrightarrow \bar{w}$. Since $\lambda$ is a soft supersymmetry breaking
parameter and $\zeta$ is the parameter breaking $N=2$ to $N=1$, the 
previous symmetry instructs us (consistently) to consider only cases where the 
supersymmetry soft breaking scale is much smaller than the scale at
which supersymmetry is broken to $N=1$. We solve the relation between the
parameters by setting $\zeta=\mu\eta$ and $\lambda = -\epsilon/\mu$. 
The $U(1)$ charge conservation and the supersymmetric limit constraint
$\eta$ to depend on the $N=2$ dynamical scale as $\eta \propto (\Lambda_2)^2
\left( 1 + O(\epsilon)\right)$ and the $N=1$ breaking parameter 
$\mu$ is related to the mass of the adjoint 
chiral multiplet by $\mu \propto m_{Adj}\left( 1 + O(\epsilon)\right)$ 
\refs\rthetus.
Similar arguments based on U(1) charge conservation and holomorphy in the 
supersymmetric limit,
fix the dependence of the punctures to be 
$z_\pm=F(N_c,N_f,r,\Lambda_2,m_f) + O(\Lambda_2 \epsilon)$
where $F$ is fixed by the comparison of the $N=2$ curve (i.e.\ the 
$\ep, \mu \rightarrow 0$ limit) with the corresponding
Seiberg-Witten curve \rmberk. 
Notice that the freedom to shift the punctures does not introduce 
new supersymmetry breaking parameters.

Notice also that supersymmetry breaking ($\epsilon\neq 0$) introduces
an angle in the $(X^6,X^7)$ plane between the NS-branes and the 
semi-infinte D4-branes, which are aligned along $X^6$. 
A possible interpretation of this angle, 
which is not an independent parameter but is proportional to 
$\sqrt{\epsilon}$, is as an induced squark mass.~\foot{We thank 
N.\ Evans for discussion on this point.}

\newsec{Run-away and decoupling of adjoint matter in softly broken curves}
In the supersymmetric case, it is well known \refs\rmberk\ that the curve 
eq.\ \ourcurve\ does not lead to run-away in the decoupling limit of
the adjoint chiral multiplet (i.e.\ $\vert\mu\vert\rightarrow\infty$)
when $N_f>N_c$ if $m_f=0$, and when $r=0$ if $m_f\neq 0$. This
is the correct behaviour known from field theory \rads. 

It is interesting to see what happens in the decoupling limit of
the adjoint chiral multiplet with soft supersymmetry breaking. We first
consider the massless case $m_f=0$. 
Recall \rmberk\ that in this case the matching between the $N=2$ and
$N=1$ scales reads 
$(\Lambda_1)^{3N_c-N_f}= (m_{Adj})^{N_c} (\Lambda_2)^{2N_c-N_f}$.
Since our curve describes soft supersymmetry breaking, the relevant
scale after we decouple the adjoint chiral multiplet is $\Lambda_1$,
while the supersymmetry breaking scale is much smaller. This
implies that in the limit $\vert\mu\vert\rightarrow\infty$ we have
$\Lambda_2 \sim \mu^{-N_c/(2N_c-N_f)}$. 

It is possible to determine the scaling of $\epsilon$ in terms of $\mu$ and 
$\Lambda_2$ by requiring that the various components of the curve have a
meaningful physical interpretation in the $\vert\mu\vert\rightarrow\infty$ 
limit.

In our notation, the world-sheet parameter, $z$, has dimensions of
energy. This means that $z$ has to be scaled as we take decoupling
limits, like $\vert\mu\vert\rightarrow \infty$. In particular, since the
softly broken curves are constructed as deformations of the $N=2$
curve, the interesting features of the curve occur at
$z\sim O(\Lambda_2)$ as long as 
$\vert\mu\vert, \vert\epsilon\vert \ll \vert\Lambda_2\vert$. However,
in the $\vert\mu\vert\rightarrow \infty$ limit, the right angle limit 
of the NS'-brane is a singular configuration where the original
curve factorizes in components. The relevant scale of the curve
now is the $N=1$ dynamical scale $\Lambda_1$, and one can easily
check that $z\sim O(\Lambda_1)$ only captures the NS component
of the degenerate $N=1$ curve.

In general, in order to isolate different structures in the
curve surviving in the $\vert\mu\vert\rightarrow \infty$ limit, we write  
 $z=\alpha u$, with $u\sim O(\Lambda_1)$. 
Now if $\alpha$ is chosen of order $O(1)$,
in the $\vert\mu\vert\rightarrow\infty$ limit only the component of 
the NS-brane at
$\vert z\vert=\infty$ (the non-rotated brane) remains; This component
is supersymmetric: $v=u$, $w=0$, $t\sim u^{N_c-N_f}$ and $X^7=0$. 
%
For $\alpha=O(\vert\mu(\Lambda_2)^2 /(\Lambda_1)^2\vert)$ the 
component of the rotated NS'-brane remains. In this case 
$v=\epsilon\vert\Lambda_1\vert^2/
[\vert\mu(\Lambda_2)^2\vert \bar{u}]$, $w\sim\vert\Lambda_1\vert^2/u$,
$t\sim (u)^{N_c}$ and $X^7=0$. From the last expression for $v$ we see that 
to stabilize this component in the $\vert\mu\vert\rightarrow\infty$ 
limit we need to 
set $\epsilon\sim \mu(\Lambda_2)^2$. More precisely we set
\eqn\epslam{\epsilon = \vert\mu(\Lambda_2)^2\vert \,
f_\epsilon (\vert m_\lambda\vert,\vert\Lambda_1\vert, R,\theta)}
where by $m_\lambda$ we indicate the effective supersymmetry breaking
mass scale, to be identified at first order with the gaugino mass. 
The real function $f_\epsilon$ must vanish for $m_\lambda=0$ and
depends also on the phases of all complex quantities which we denoted
collectively by $\theta$; this dependence is constrained by the $U(1)$ 
symmetries. 

In the massless case ($m_f=0$), the punctures scale as 
$z_\pm =\Lambda_2 \left( c_\pm + O(\epsilon)\right)$.~\foot{Using
$SL(2)$ invariance, we can keep fixed the $z=0$ and $z=\infty$ punctures,
and allow a shift of $z_+$ and $z_-$ by a common factor proportional to 
$\epsilon$.}
{}From the scaling of $\epsilon$ determined in eq.\ \epslam\ we see that
as long as $N_f < 2N_c$, the non-supersymmetric shift of the punctures 
is irrelevant in the $\vert\mu\vert\rightarrow\infty$ limit  since 
$\mu(\Lambda_2)^2$ vanishes.

Let us now consider the issue of run-away. For $r\neq 0$ we should look if
$w\vert_{z_\pm}$ and $v\vert_{z_\pm}$ run-away or remain finite in the
$\mu\rightarrow\infty$ limit.  For $r=0$ only the puncture $z_-$ is to be
considered. The conditions to stop the run-away are
\eqn\runstop{ v\vert_{z_\pm} \sim {\epsilon\over\Lambda_2} < \infty \ ,
\qquad\qquad\quad 
w\vert_{z_\pm} \sim \mu\Lambda_2 -{\epsilon\over \bar\mu\bar\Lambda_2}
< \infty }
in the  $\vert\mu\vert\rightarrow\infty$ limit. From the first condition
it is evident that only if $\epsilon/\Lambda_2$ is kept finite in the 
limit, then there is 
not run-away induced by the supersymmetry breaking terms. But
if $\epsilon/\Lambda_2$ is finite, the second equation in \runstop\ implies
that $w\vert_{z_\pm}$ has the same limit as in the supersymmetric case.
It follows that run-away can be stopped {\sl only\/} in the cases in 
which it is also stopped in absence of soft supersymmetry breaking. Moreover
$\epsilon$ should vanish at least as fast as $\Lambda_2$ in this limit,
which is already guaranteed by eq.\ \epslam. 

As discussed in the introduction, the fact that the soft supersymmetry 
breaking encoded in the curves is not able to stop the 
run-away, implies that the squark mass generated radiatevely by a tree
level gaugino mass $m_\lambda$ in ordinary softly broken SQCD, 
is not properly captured by these curves. 
Thus for $N_f < N_c$ MQCD fails to reproduce
the vacuum structure of softly broken SQCD.

To summarize, in the limit $\vert\mu\vert\rightarrow\infty$, the NS-brane 
 component of the curve is supersymmetric, 
whereas the component of the rotated NS'-brane, is not:
\eqn\fincurv{\eqalign{& \alpha \sim O(1)\,:\qquad\qquad\quad v=u \ ,\qquad
w=0\ ,\qquad t\sim u^{N_c-N_f}\ ,\qquad X^7=0\cr
& \alpha \sim O\left(\left\vert{\mu(\Lambda_2)^2\over \Lambda_1}\right\vert
\right)
\,:\quad v={\vert\Lambda_1\vert^2 f_\epsilon\over \bar{u}} \ ,\quad
w={\vert\Lambda_1\vert^2c_+'c_-'\over u} \ ,
\quad t\sim u^{N_c}\ ,\quad X^7=0\ .\cr }}
For $N_f > N_c$, we can also identify  a third component containing
the flavour D4-branes, degenerated in the $v=w=0$ plane:
\eqn\finc{
 \alpha \sim O\left(\left\vert{\Lambda_2\over \Lambda_1}\right\vert
\right)\,:\qquad ~~ v=0 \ ,\qquad
w=0\ ,\quad t\sim {u^{N_c-N_f}\over (u-c_+')^r (u-c_-')^{N_f-r}}
\ ,\quad X^7=0.} 
This component merges with the NS'-brane component in the $N_c = N_f$
case, since then we have $\mu (\Lambda_2)^2 \sim \Lambda_2$.   
This is of course what we expected 
to happen from the discussion in the previous section. 

In the case $m_f\neq 0$ and $r=0$ (see also ref.\ \rlastev), 
the analysis is much simpler since
it turns out that the curve remains in a single component and $\epsilon$
is constant in the $\vert\mu\vert\rightarrow\infty$ limit. Thus we have
$\epsilon=g_\epsilon(\vert m_\lambda\vert,\vert m_f\vert,\vert\Lambda_1\vert
,R,\theta)$, where $g_\epsilon$
vanishes if either of $m_\lambda$ or $m_f$ vanishes. 
The curve is given by eq.\ \ourcurve\ with $r=0$, $\lambda=0=\eta$, 
$\zeta=-c_+ m_f$, $z_+=0$, $z_-=-m_f$ and 
$\vert\Lambda_2\vert\rightarrow\vert\Lambda_1\vert$.

It is interesting to notice that expanding 
$\epsilon(m_\lambda,m_f,m_{Adj},\Lambda_1,R)$
in positive powers of $m_\lambda$ for very small $m_\lambda$, and 
in the $\vert m_{Adj}\vert\rightarrow\infty$ limit, the $U(1)$ charges 
imply that the most general form for $\epsilon$ is~\foot{Recall that 
$m_{Adj} (\Lambda_2)^2\rightarrow 0$  in the 
$\vert m_{Adj}\vert\rightarrow\infty$ limit.}
\eqn\geneps{\epsilon = m_\lambda R^2 \left\{ \left[ (\Lambda_1)^{3N_c-N_f} 
(m_f)^{N_f} \right]^{1\over N_c} + D\, m_{Adj} (\Lambda_2)^2 \right\} \ +\ 
{\rm c.c.}\ +\ \hbox{\rm higher orders} }
where $D$ is an adimensional constant. This gives the explicit form
of $f_\epsilon$ and $g_\epsilon$ to first order in $m_\lambda$.

\newsec{Acknowledgements}
We thank J.\ Sonnenschein for sharing his insights with us and
for comments on the manuscript. We also thank N. Evans for discussions.

This work is partially supported by the European Commission TMR programme
ERBFMRX-CT96-0045 in which A.P.\ is associated to the Institute for 
Theoretical Physics, K.U.\ Leuven. A.P.\ would like to thank CERN for its
hospitality while part of this work was carried out.

\listrefs
 \vfill\eject
 \bye
\end